\documentclass[11pt]{article}

\usepackage[T1]{fontenc}
\usepackage[utf8]{inputenc}
\usepackage{lmodern}
\usepackage[margin=1in]{geometry}
\usepackage{amsmath,amssymb}
\usepackage{graphicx}
\usepackage{booktabs}
\usepackage{tabularx}
\usepackage{array}
\usepackage{makecell}
\usepackage[dvipsnames]{xcolor}
\usepackage[labelfont=bf,labelsep=period,font=small]{caption}
\usepackage[separate-uncertainty=true,per-mode=symbol]{siunitx}
\usepackage[numbers,sort&compress,square]{natbib}
\usepackage{authblk}
\usepackage{microtype}
\usepackage{xurl}
\usepackage[colorlinks=true,linkcolor=blue!60!black,citecolor=blue!60!black,urlcolor=blue!60!black]{hyperref}
\usepackage{doi}

\graphicspath{{figures/}}
\DeclareSIUnit{\RIU}{RIU}
\newcolumntype{L}{>{\raggedright\arraybackslash}X}

\newcommand{\figref}[1]{Fig.~\ref{#1}}
\newcommand{\figsref}[1]{Figs.~\ref{#1}}

\title{Laser-direct-written polymer WGM microring laser integrated on a fiber tip for refractive index sensing}

\author[1,\textdagger]{Mobin Yaghoobi}
\author[1,\textdagger]{Seyedeh Mahshad Hosseini}
\author[1]{Leila Hajshahvaladi}
\author[1,*]{Gholam-Mohammad Parsanasab}
\author[2]{Mohammad Ismail Zibaii}

\affil[1]{Integrated Photonics Laboratory, Faculty of Electrical Engineering, Shahid Beheshti University, Tehran, Iran}
\affil[2]{Laser and Plasma Research Institute, Shahid Beheshti University, Tehran, Iran}

\date{}

\begin{document}
\maketitle
{\renewcommand{\thefootnote}{}%
 \footnotetext{\textsuperscript{\textdagger}These authors contributed equally to this work.\\
 \textsuperscript{*}Corresponding author: \href{mailto:gm_parsanasab@sbu.ac.ir}{gm\_parsanasab@sbu.ac.ir}}}

\begin{abstract}
Whispering-gallery-mode (WGM) microlasers are promising for refractometric sensing, but their practical implementation can be limited by the need for stable resonator--coupler alignment. Here, we demonstrate a Rhodamine B-doped polymer WGM microring laser directly fabricated on an optical-fiber tip, eliminating the external resonator--fiber coupling interface at the sensing head. Micro-droplet coating and UV laser direct writing are combined for fabrication, while finite-difference time-domain simulations guide the microring laser design. The microlasers operate in air and water with thresholds of \SI{0.224}{\micro\joule} and \SI{0.458}{\micro\joule}, respectively. Refractometric measurements in aqueous glucose solutions yield a bulk refractive index sensitivity of \SI{364.95}{\nano\metre\per\RIU} and a detection limit of \SI{2.47e-4}{\RIU}. Water-induced polymer changes and photobleaching are also investigated. This fiber-tip architecture provides a basis for probe-based refractometric sensing in environmental and biochemical applications.
\end{abstract}

\section{Introduction}

Changes in a medium's composition or concentration can alter its refractive index (RI), providing a basis for chemical and biological sensing. Optical whispering-gallery-mode (WGM) resonators have shown promising performance in such sensing applications because of their high quality ($Q$) factors and enhanced evanescent-field interaction with the surrounding medium~\cite{Yu2021}. Optical gain enables active WGM resonators to produce narrow-linewidth lasing modes~\cite{Reynolds2017}. A theoretical comparison of active and passive microspheres reported a $10^{4}$-fold linewidth reduction after introducing optical gain~\cite{Yang2006}. Experimentally, an erbium-doped microsphere laser showed higher RI sensitivity and a lower detection limit compared with its passive counterpart~\cite{Yan2021}. This highlights the potential of active resonators to resolve small spectral shifts resulting from variations in the modal effective refractive index~\cite{Capocefalo2023,Krammer2017}. Active WGM resonators also allow excitation and readout through either free-space configurations or external coupling elements~\cite{Toropov2021}. The coupling scheme can influence the lasing characteristics, sensing performance, and practicality of active WGM sensors. Although free-space configurations have enabled applications such as intracellular sensing and live-cell tracking~\cite{Toropov2021}, their use can become challenging over longer optical paths, particularly in scattering or absorbing media~\cite{Ni2026}. Alternatively, evanescent coupling through prisms, waveguides, or tapered fibers enables efficient WGM excitation and collection~\cite{Gorajoobi2019}. In dye-doped polymer microdisk lasers, for example, evanescent fiber excitation reduced the lasing threshold by approximately two orders of magnitude compared with free-space pumping~\cite{Siegle2018}.

Efficient evanescent coupling requires proper modal overlap and phase matching, and the coupling strength is sensitive to the resonator--coupler gap~\cite{Gorajoobi2019}. In an active microring--waveguide configuration, varying this gap can affect the coupling loss, $Q$ factor, lasing threshold, and number of lasing modes~\cite{Arvanagh2025}. Maintaining stable coupling therefore requires precise positioning, mechanical stability~\cite{Junge2011}, and protection of the resonator--coupler interface from contamination~\cite{Monifi2013}, particularly under non-laboratory conditions. To improve robustness and long-term coupling stability, microfluidic integration~\cite{Wang2020} and low-index polymer packaging~\cite{Monifi2013} have been explored; however, these approaches can increase implementation complexity and the sensing-head footprint, limiting their practicality in confined or difficult-to-access environments~\cite{Liu2025}. Alternatively, integrating WGM microresonators with optical fibers has been explored to realize compact sensing probes with fiber-mediated excitation and readout~\cite{Yang2023}. One approach incorporates prefabricated microresonators into fiber structures, such as a dye-doped polymer microsphere positioned at a microstructured fiber tip~\cite{Francois2015} and an Er-doped microsphere placed in a micromachined cavity within a standard fiber~\cite{Yan2021}. Xu \emph{et al.} transferred a polymeric membrane carrying a microring laser onto a flat fiber tip for humidity sensing~\cite{Xu2021}. In such implementations, the separately fabricated microresonator needs to be properly assembled with the fiber structure and maintained in a stable position. Another approach is to use the fiber itself as the fabrication substrate to create the microresonator \emph{in situ}, avoiding the subsequent transfer and assembly~\cite{Xiong2020,Halendy2023}.

Direct site-selective fabrication on optical fibers is challenging due to their geometry and small cross-section~\cite{Xiong2020,Kostovski2014}. Femtosecond-laser micromachining has been used to fabricate cylindrical~\cite{Shi2015} and microsphere~\cite{Zhao2026} cavities in fibers, but surface roughness can limit the $Q$ factor of laser-machined cavities~\cite{Shi2015}. Laser direct writing (LDW) of photopolymers enables maskless micro- and nanofabrication on optical fibers~\cite{Wu2019,Xiong2020b}. Using femtosecond two-photon polymerization, polymer WGM resonators with nanometer-scale sidewall roughness have been fabricated~\cite{Tomazio2017}. Femtosecond LDW has also been used to fabricate 3D fiber-tip sensing structures in which passive WGM resonators are integrated with elements that couple light between the fiber and the resonator~\cite{Halendy2023,Liu2020,Nan2025}. Active WGM resonators can eliminate the need for such integrated coupling structures~\cite{Reynolds2017}, reducing the fiber-tip architecture to the resonator itself. Planar resonators such as microrings and microdisks can therefore be patterned by one-photon LDW using continuous-wave lasers, avoiding a femtosecond writing laser and potentially reducing fabrication cost and setup complexity~\cite{Wu2019,Do2014}. In addition, polymers can incorporate diverse optical gain materials, offering versatility in active WGM resonator design~\cite{Reynolds2017,Toropov2021}. The photostability of the gain medium, however, should be considered during optical excitation, as photobleaching and related photoinduced changes can affect the laser output~\cite{Vellaichamy2023}. For example, Pyrromethene-597-doped PMMA WGM microdisk lasers showed decreasing lasing intensity and a blue shift attributed to photobleaching~\cite{Wondimu2018}. In aqueous sensing, water uptake by the polymer can also contribute to shifts in the WGM resonances~\cite{Saunders2015,Eryurek2017}. Fabricating polymer microresonators directly on fiber tips also presents challenges in achieving adequate polymer adhesion to the silica facet, forming a uniform polymer layer for lithography, and mitigating polymerization-induced shrinkage. These challenges, together with the optical and material limitations discussed above, need to be addressed when developing practical fiber-tip WGM sensing probes.

In this work, we demonstrate a Rhodamine B-doped polymer WGM microring laser directly fabricated on an optical-fiber facet and aligned with the fiber core for refractometric sensing. The microring is fabricated by combining micro-droplet coating with LDW lithography. Finite-difference time-domain (FDTD) simulations are used to investigate the effect of ring radius on the resonance spectrum and bulk refractive index sensitivity (BRIS), guiding the selection of the ring geometry for fabrication. We experimentally characterize the microlaser in air and water and evaluate its spectral response to the refractive-index change of the glucose solutions. The effects of photobleaching and water-induced polymer changes on the microlaser operation are also investigated. In contrast to chip-based WGM sensors that often require microfluidic integration and additional packaging for operation in liquid environments, this active fiber-tip architecture eliminates the external resonator--fiber coupling interface at the sensing head and its associated requirements for coupling-gap control and mechanical stability. The resulting compact sensing head provides a basis for the development of directly immersible microlaser probes for \emph{in situ} sensing, with potential for future \emph{in vivo} applications.

\section{Simulation and design principle of the fiber-tip microring laser sensor}
\label{sec:design}

\begin{figure}[tb]
  \centering
  \includegraphics[width=0.78\linewidth]{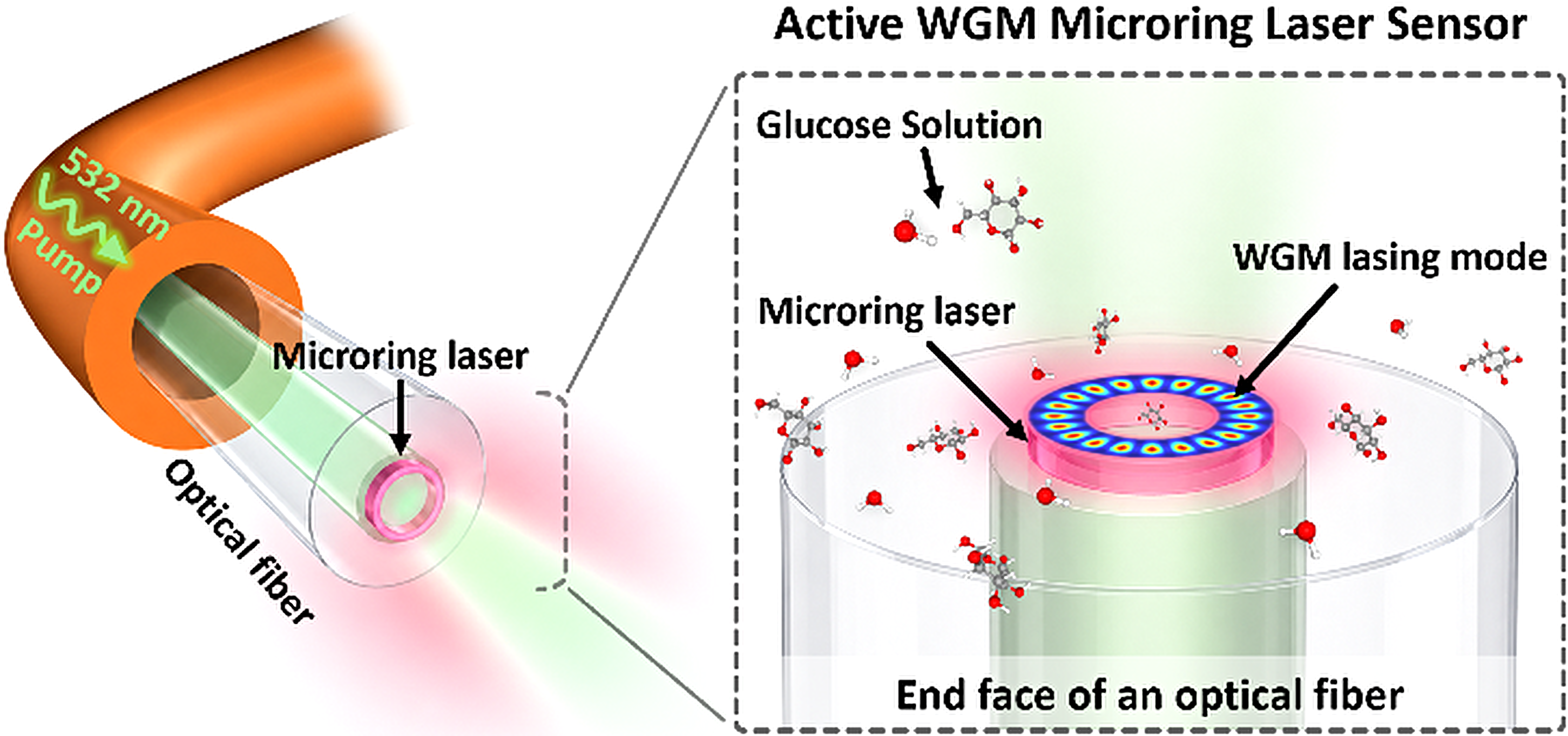}
  \caption{Conceptual illustration of the fiber-tip WGM microring laser refractometric sensor, showing the interaction of the microring laser evanescent field with the surrounding medium.}
  \label{fig:concept}
\end{figure}

This section presents the FDTD simulations of the refractive-index sensing mechanism in active WGM microrings and the design considerations of the proposed fiber-tip microring laser sensor. An illustration of the proposed fiber-tip microring laser for bulk refractive-index sensing is depicted in \figref{fig:concept}. Among the sensing mechanisms explored for active WGM microresonators, tracking the spectral position of the lasing modes is commonly used to detect changes in the bulk refractive index~\cite{Reynolds2017}. The resonance wavelength of a WGM microlaser depends on the cavity geometry and the modal effective refractive index. A change in the surrounding refractive index modifies the effective optical path length of the lasing WGM and shifts its resonance wavelength. The magnitude of this response depends on the interaction of the optical mode with the surrounding medium, which is examined below through the simulated microring geometry.

\begin{figure}[tb]
  \centering
  \includegraphics[width=\linewidth]{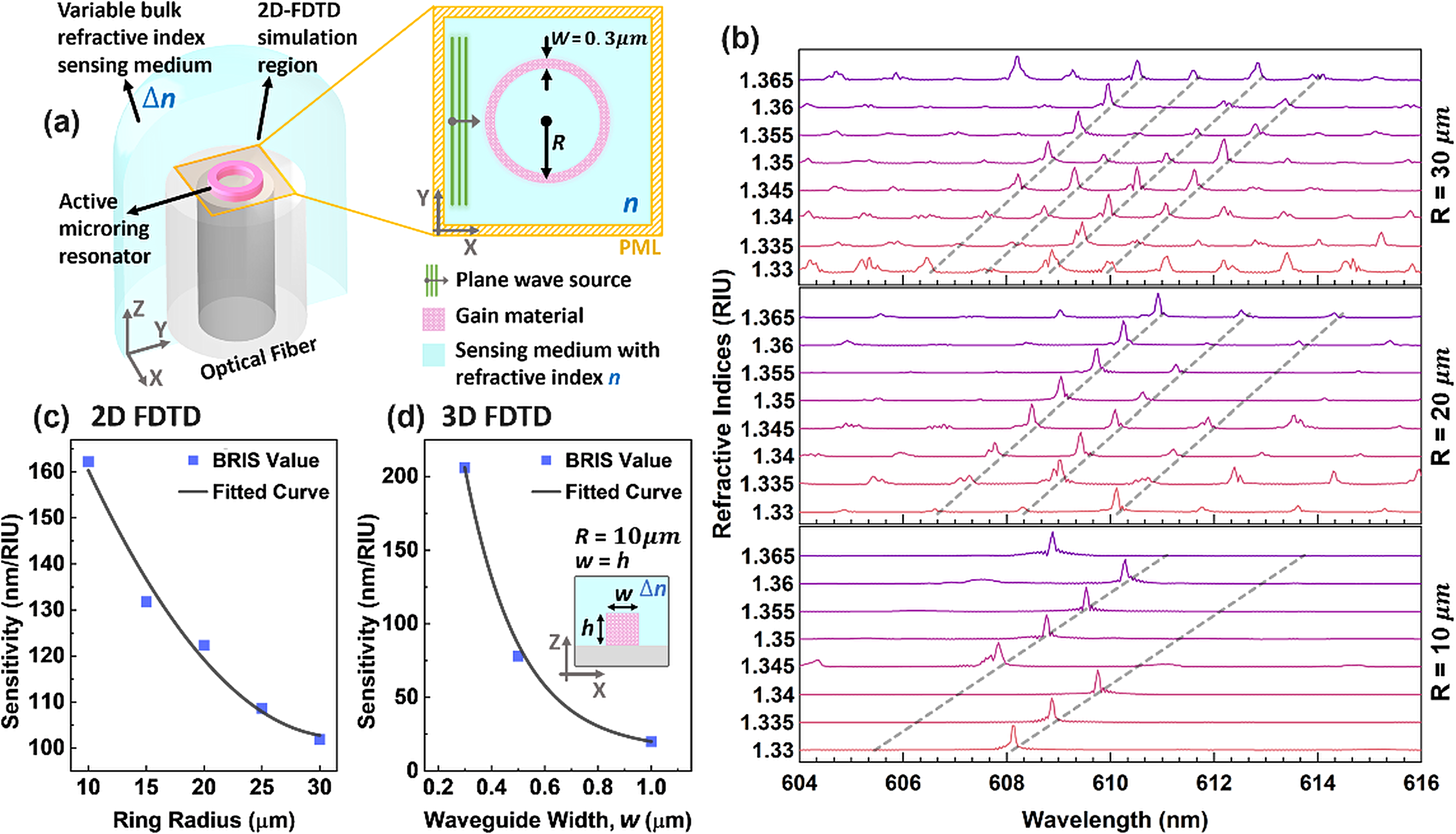}
  \caption{FDTD simulation of the refractive-index sensing performance of active WGM microrings. (a)~Simulated 2D microring with radius $R$, waveguide width $w$, and surrounding RI $n$. (b)~Normalized emission spectra for radii of 10, 20, and \SI{30}{\micro\metre} over an RI range of 1.33 to 1.365, showing redshifts and a single- to multimode transition. (c)~Calculated BRIS versus ring radius. (d)~Calculated BRIS versus waveguide cross-sectional dimensions using 3D-FDTD with approximately \SI{10}{\nano\metre} sidewall roughness.}
  \label{fig:sim}
\end{figure}

We performed FDTD simulations incorporating a frequency-dependent gain medium to emulate the sensing performance of an active cavity. The primary objective was not to determine the exact microring dimensions for fabrication, but rather to investigate the influence of key geometrical parameters on refractive-index sensitivity. As illustrated in \figsref{fig:sim}(a) and \ref{fig:sim}(b), a 2D model was used to evaluate the effect of microring radius on the sensing response. Because the resonance-wavelength shift is primarily governed by the interaction between the microring's evanescent field and the surrounding medium, this model provides a reasonable estimate of how the sensitivity varies with microring radius. Microrings with radii from \SIrange{10}{30}{\micro\metre} were investigated using a fixed waveguide width of \SI{0.3}{\micro\metre}. The microring laser was laterally pumped with a TE-polarized plane-wave source at \SI{532}{\nano\metre}, and perfectly matched layer (PML) boundary conditions were applied to suppress artificial reflections. The refractive index of the surrounding medium was varied from 1.33 to 1.365, corresponding to the RI range of the aqueous glucose solutions.

\figref{fig:sim}(b) shows the simulated emission spectra for representative ring radii of 10, 20, and \SI{30}{\micro\metre}. Increasing the surrounding refractive index produces an overall redshift of the lasing resonances. The \SI{10}{\micro\metre} ring exhibits predominantly single-mode behavior within the simulated gain bandwidth, whereas larger rings support more lasing modes because their smaller free spectral range allows additional cavity modes to fall within the gain bandwidth~\cite{Sorayaie2025}. At some RI values, the dominant lasing peak shifts discontinuously to a different cavity mode, resulting in an apparent shift toward shorter wavelengths. This behavior is consistent with mode hopping between competing lasing modes and does not indicate a reversal of the underlying refractive-index response.

\figref{fig:sim}(c) shows the simulated refractive-index sensitivity as a function of the microring radius. The sensitivity increases as the ring radius decreases. The resonance condition of a microring is given by $m\lambda_{\mathrm{res}} = n_{\mathrm{eff}}\,(2\pi R)$, where $m$ is the azimuthal mode number and $n_{\mathrm{eff}}$ is the effective refractive index of the guided mode. For a fixed ring radius and mode number, changes in the surrounding refractive index $n$ modify $n_{\mathrm{eff}}$ and consequently shift the resonance wavelength. Considering the wavelength dispersion of the effective refractive index, the resonance-wavelength sensitivity can be expressed as
\begin{equation}
  S_{\mathrm{RI}} = \frac{d\lambda_{\mathrm{res}}}{dn} = \frac{\lambda_{\mathrm{res}}}{n_{g}}\,\frac{\partial n_{\mathrm{eff}}}{\partial n},
  \label{eq:sensitivity}
\end{equation}
where $n_{g} = n_{\mathrm{eff}} - \lambda_{\mathrm{res}}\,(\partial n_{\mathrm{eff}}/\partial\lambda)$ is the group index. Equation~\eqref{eq:sensitivity} shows that the sensitivity depends on how strongly $n_{\mathrm{eff}}$ responds to changes in the surrounding RI. Greater evanescent-field overlap increases this response and therefore enhances $S_{\mathrm{RI}}$. For the simulated microrings, reducing the radius increases the field interaction with the surrounding medium because of the stronger curvature and weaker modal confinement, accounting for the trend in \figref{fig:sim}(c).

\figref{fig:sim}(d) further presents the effect of the waveguide cross-sectional dimensions on the refractive-index sensitivity. The 3D-FDTD simulations were performed for a microring with a radius of \SI{10}{\micro\metre} and approximately \SI{10}{\nano\metre} sidewall roughness, while the waveguide width and thickness were varied. The sensitivity decreases with increasing waveguide cross-sectional area because stronger modal confinement reduces the evanescent-field overlap with the surrounding medium. Thus, within the investigated range, smaller waveguide cross sections provide higher refractive-index sensitivity.

The simulated sensitivity trends alone do not determine the optimum geometry for the fabricated fiber-tip microlaser. In practice, the performance of fabricated polymer microring lasers can deviate from that predicted for ideal structures because fabrication-induced imperfections, such as surface nonuniformity and sidewall roughness, introduce additional optical losses. These effects can be mitigated through material-compatible pre- and post-fabrication treatments and high-resolution lithographic techniques such as laser direct writing. Nevertheless, selecting an appropriate ring radius for a fiber-tip microlaser sensor requires a fabrication-aware design approach that accounts for both fabrication-related constraints and the intrinsic optical properties of the resonator. To this end, we adopted the analytical design approach previously established in Ref.~\cite{Sorayaie2025}, which provides a tradeoff among the geometry of the fabricated polymer microrings, surface-scattering and bending losses, and the lasing threshold. Although reducing the ring radius can theoretically improve the refractive-index sensitivity, radiation and scattering losses also increase at smaller radii, which can suppress lasing operation. Sidewall roughness also introduces additional scattering loss, which is a critical factor in the optimal WGM microcavity design. A reduction in the radius enhances the electric field intensity at the waveguide sidewalls, leading to increased scattering and a reduced $Q$ factor. Smaller rings also exhibit higher lasing thresholds, requiring higher pump powers that may induce thermal damage, polymer degradation, or photobleaching of the dye molecules~\cite{Vellaichamy2023}. Maintaining a low lasing threshold is therefore important for both energy-efficient operation and long-term stability of the dye-doped polymer laser. Besides these constraints, the finite core area of the optical fiber imposes an additional limitation. To ensure efficient optical pumping, the microring should remain within the \SI{62.5}{\micro\metre}-diameter core of the multimode fiber used in this work, where the pump intensity is highest. Therefore, a ring radius of \SI{20}{\micro\metre} was selected for experimental realization as a practical compromise between sensing performance, lasing efficiency, and compatibility with the fiber-tip geometry.

\section{Fabrication and experimental characterization methods}
\label{sec:methods}

\begin{figure}[tb]
  \centering
  \includegraphics[width=\linewidth]{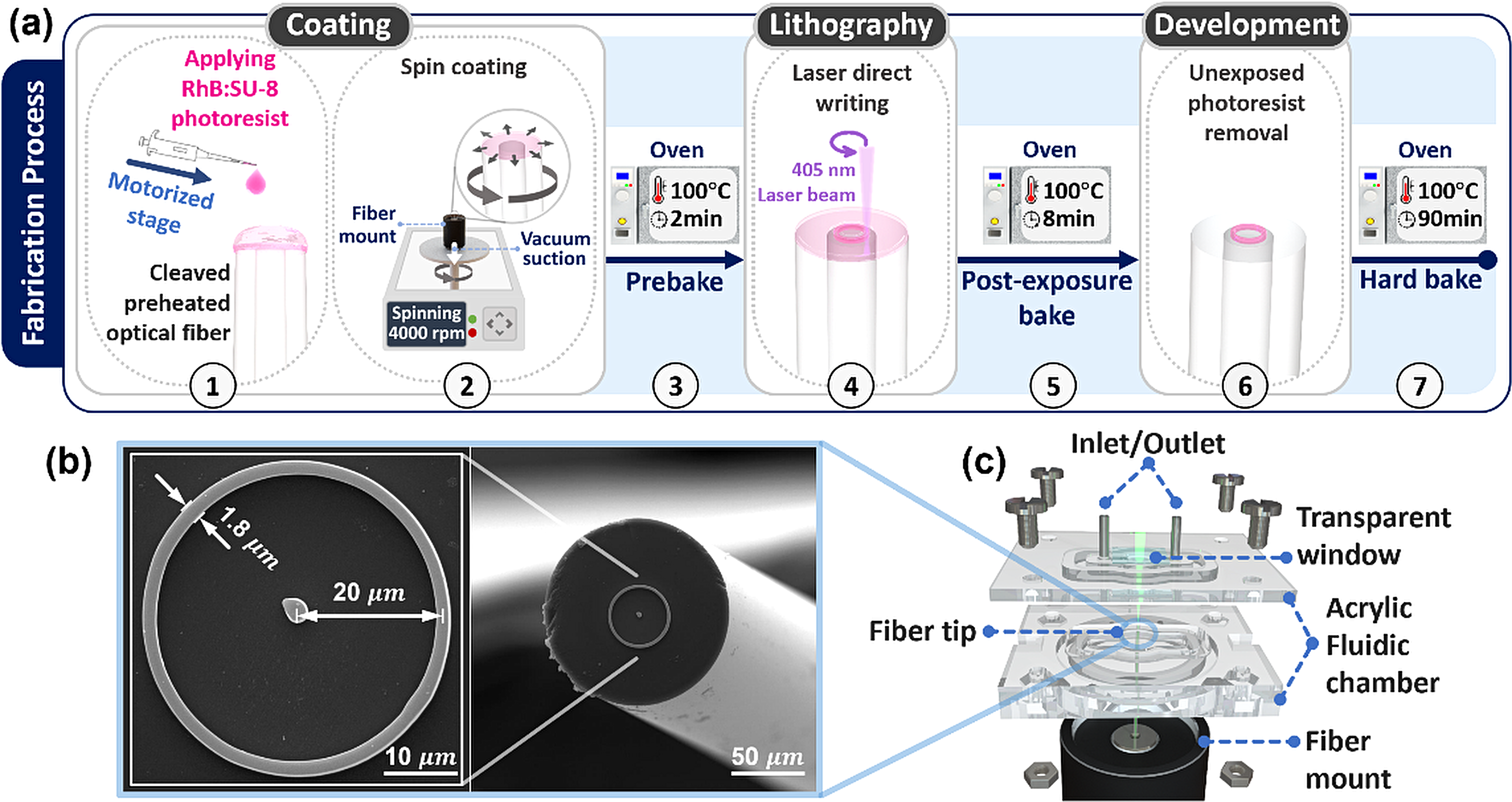}
  \caption{Fabrication and structural characterization of the fiber-tip microring laser. (a)~Fabrication steps: RhB:SU-8 coating, laser direct writing, and development, each followed by a baking step. (b)~SEM image of a fabricated microring laser on a flat fiber tip with $R=\SI{20}{\micro\metre}$ and $w=\SI{1.8}{\micro\metre}$. (c)~Schematic of the custom-designed acrylic fluidic chamber for refractometric sensing and optical characterization in aqueous environments.}
  \label{fig:fab}
\end{figure}

The fabrication of active polymeric microring lasers on the fiber tip was carried out through a multi-step process combining localized droplet coating, LDW lithography, and photoresist development, as schematically illustrated in \figref{fig:fab}(a). A multimode optical fiber (MMF-OM1) was cleaved to obtain a near-perpendicular end face (cleave angle $<\ang{1}$), verified using the inspection module of a fusion splicer. The cleaved fiber was inserted into a standard FC/PC connector, with approximately \SI{1}{\centi\metre} of the fiber protruding from the connector. The connector was mounted on a custom holder to ensure that the fiber axis was perpendicular to the subsequent coating and lithography planes. Prior to coating, the fiber tip was cleaned with isopropanol under an optical microscope and heated at \SI{100}{\celsius} for \SI{2}{min} to remove residual moisture and enhance photoresist adhesion. The active polymer solution was prepared by dissolving \SI{0.6}{\milli\gram} of Rhodamine B (RhB) in \SI{0.6}{\gram} of SU-8 photoresist (SU-8 2002, MicroChem) and diluting the mixture with \SI{2}{\milli\litre} of gamma-butyrolactone. A small droplet of the RhB:SU-8 solution was deposited on the fiber tip using a micropipette mounted on a motorized stage, forming a hemispherical meniscus. The fiber was then spin-coated at \SI{4000}{rpm} under vacuum fixation to form a uniform polymer layer on the facet, followed by a soft bake at \SI{100}{\celsius} for \SI{2}{min} to evaporate the solvent and stabilize the film thickness.

Microring patterning was performed using LDW lithography. The coated fiber was positioned on a computer-controlled translation stage, and a \SI{405}{\nano\metre} laser beam was focused onto the polymer surface using a $50\times$ objective lens. The center of the fiber core was identified using the integrated imaging system, and a circular exposure path approximately concentric with the core was patterned. A post-exposure bake was then applied at \SI{100}{\celsius} for \SI{8}{min} to complete the cross-linking process in the exposed regions. The unexposed photoresist was removed by immersing the fiber tip in methoxy-2-propanol acetate developer for \SI{10}{\second}, leaving behind the polymerized microring structure. A final hard bake at \SI{100}{\celsius} for \SI{90}{min} was performed to improve the mechanical stability, adhesion, and long-term robustness of the fabricated structure. A scanning electron microscope (SEM) image of a fabricated fiber-tip microring with a radius of \SI{20}{\micro\metre} and a width of \SI{1.8}{\micro\metre} is presented in \figref{fig:fab}(b). The conical feature at the center of the ring resulted from the fiber-core alignment procedure during lithography and does not affect device performance. The ring diameter and waveguide width were primarily controlled by the LDW exposure trajectory and laser power, while the ring thickness was governed by the spin-coating parameters and photoresist viscosity. This process enabled controlled fabrication of microring structures with similar geometries across multiple samples fabricated under identical environmental and thermal conditions. In our previous work on LDW-fabricated polymer microring lasers using the same material system, we achieved a fabrication accuracy of approximately \SI{100}{\nano\metre} and sidewall roughness of \SIrange{5}{10}{\nano\metre}~\cite{Sorayaie2025}. For subsequent optical characterization and BRIS measurements in aqueous environments, each fabricated probe was inserted into a custom-designed acrylic fluidic chamber, as shown in \figref{fig:fab}(c), through a sealed opening and fixed in place to minimize mechanical drift and thermal fluctuations during experiments.

\begin{figure}[tb]
  \centering
  \includegraphics[width=\linewidth]{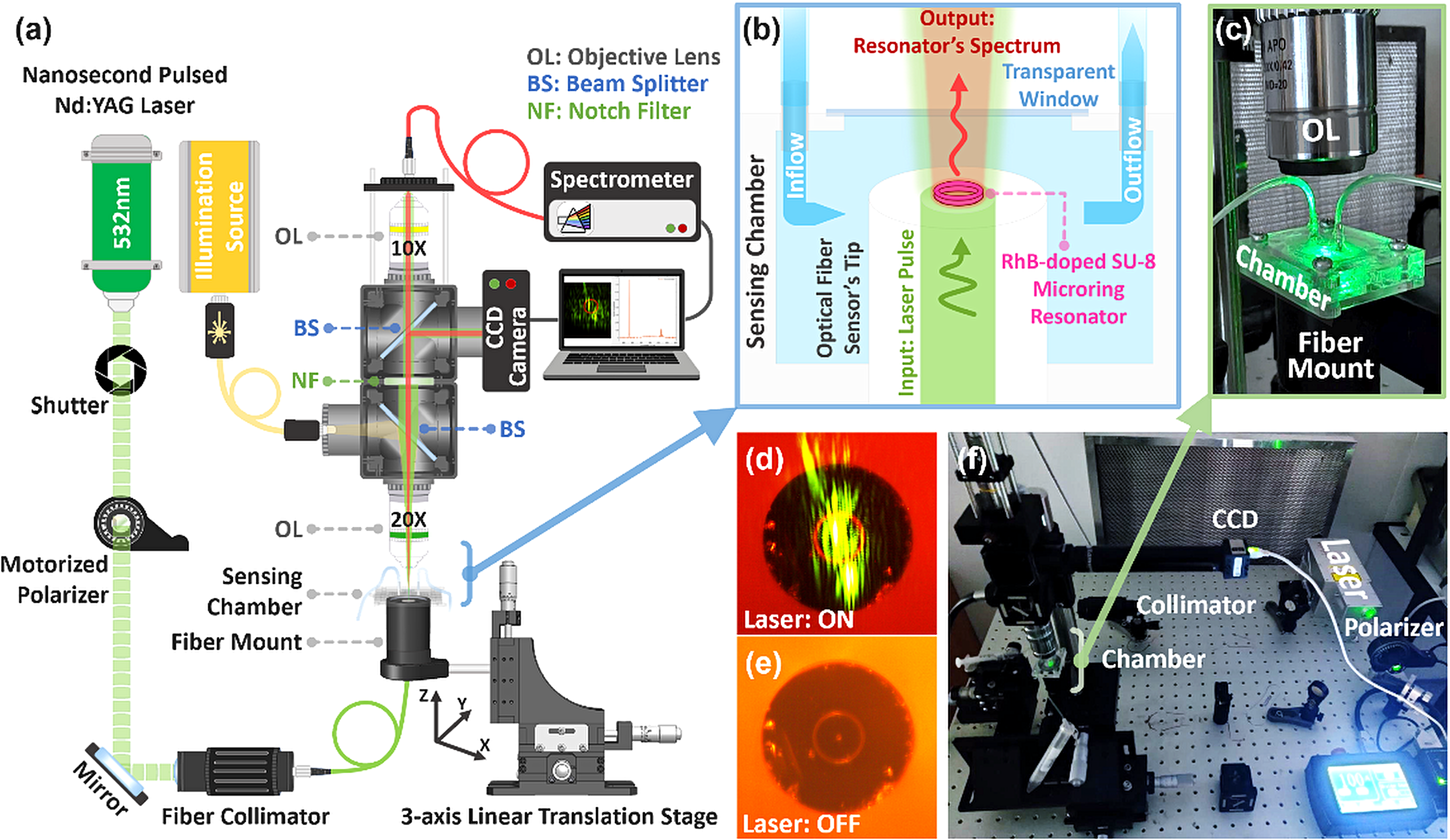}
  \caption{Optical characterization and BRIS measurement setup. (a)~Schematic of the excitation and detection system, including the \SI{532}{\nano\metre} pulsed laser, fiber coupling, collection optics, CCD camera, and spectrometer. (b)~Cross-sectional schematic of the sensing chamber, showing optical pumping through the fiber and emission collection through a transparent window. (c)~View of the sensing chamber with the mounted fiber probe. (d,~e)~CCD images of the fiber-tip microring laser with the pump laser ON and OFF, respectively. (f)~Experimental arrangement of the optical characterization setup.}
  \label{fig:setup}
\end{figure}

\figref{fig:setup} depicts the experimental setup used for optical characterization and BRIS measurements of the fabricated fiber-tip microlasers. A nanosecond pulsed Nd:YAG laser (DPS-532-E, CNI Laser) operating at a wavelength of \SI{532}{\nano\metre} was employed as the optical pump source. The laser delivered pulses with a nominal pulse duration of $<\SI{4}{\nano\second}$ and a repetition rate of \SI{10}{\hertz}. The pump energy was adjusted using a motorized polarizer and shutter, providing a controllable pump power range sufficient to exceed the lasing threshold of the RhB-doped polymer microring while avoiding photodegradation of the gain medium. The pulsed pump beam was coupled into the optical fiber through an adjustable fiber collimator, as illustrated in \figref{fig:setup}(a). The opposite end of the fiber, carrying the fabricated microring laser, was sealed inside the chamber mounted on a three-axis linear translation stage to ensure precise spatial alignment. This fiber-based pumping approach enabled efficient and stable delivery of excitation light to the active microring through the fiber core, eliminating the need for precise excitation alignments at the probe end. The emission from the microring laser was collected normal to the fiber tip using a $20\times$ objective lens. The collected light was directed toward the detection path, where a beam splitter separated the signal into two channels. One portion of the emission was sent to a CCD camera for real-time visualization and monitoring of the lasing behavior, while the remaining portion was routed to a spectrometer for spectral analysis. A narrowband optical notch filter centered at \SI{532}{\nano\metre} was placed before the spectrometer to suppress residual pump light and prevent detector saturation.

The bulk RI characterization was performed by filling the sensing chamber with aqueous solutions of known RI. The chamber was equipped with inlet and outlet ports to allow controlled fluid exchange while maintaining an isolated environment around the fiber-tip probe, as illustrated in \figsref{fig:setup}(b) and \ref{fig:setup}(c). The chamber also had a transparent window on its top side for optical access. \figsref{fig:setup}(d) and \ref{fig:setup}(e) depict representative CCD images of the fiber-tip microlaser under optical pumping (ON) and without pumping (OFF), respectively, confirming localized laser emission from the ring resonator. Arrangements of the complete experimental setup and the sensing chamber are shown in \figsref{fig:setup}(f) and \ref{fig:setup}(c), respectively.

\section{Lasing performance and refractometric sensing}
\label{sec:results}

\begin{figure}[tb]
  \centering
  \includegraphics[width=\linewidth]{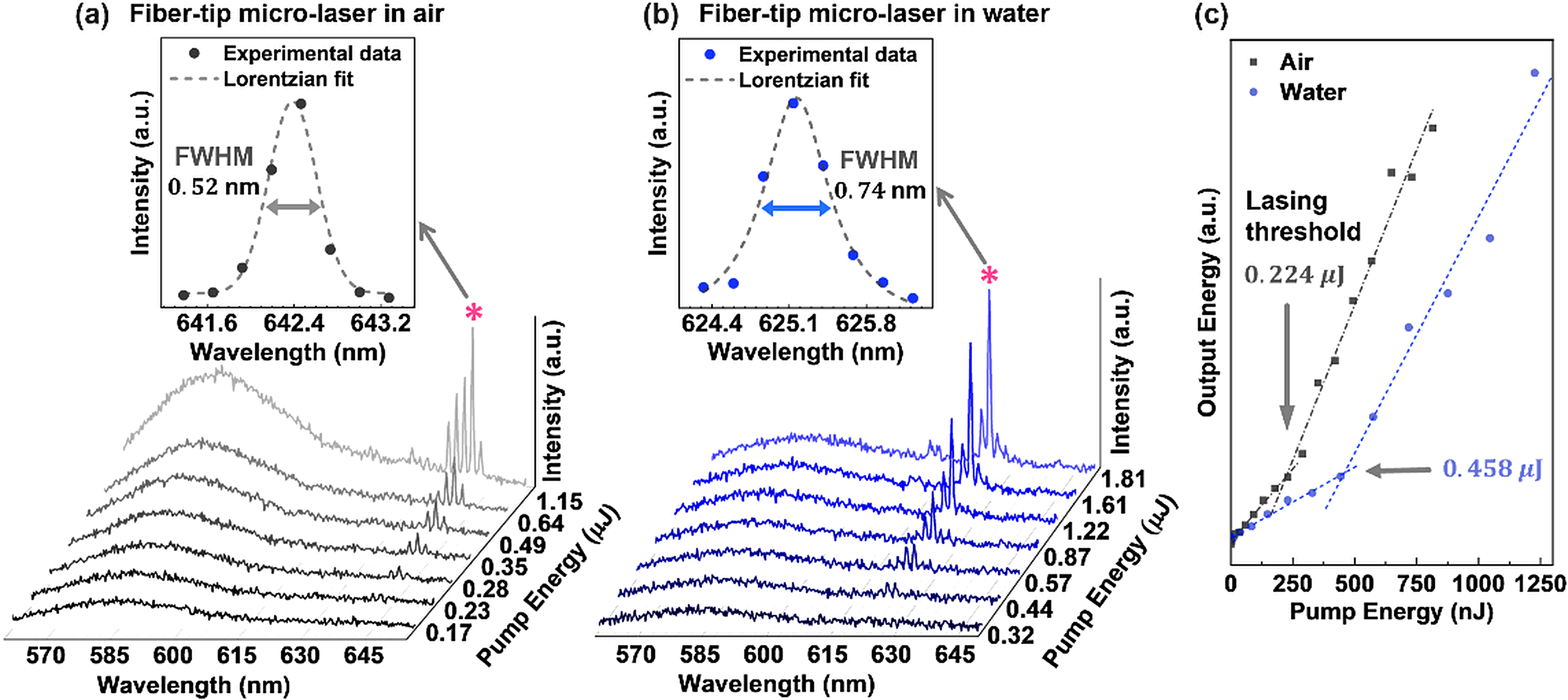}
  \caption{Lasing characteristics of two separate fiber-tip microring lasers, one characterized in air and the other in water. (a,~b)~Emission spectra versus pump pulse energy in air and water, respectively. Insets show enlarged lasing peaks and corresponding FWHM values. (c)~Output energy versus pump pulse energy, with linear fits used to determine the lasing thresholds.}
  \label{fig:lasing}
\end{figure}

Lasing characteristics were evaluated using two separate fiber-tip microring lasers, one measured in air and the other in water. \figsref{fig:lasing}(a) and \ref{fig:lasing}(b) depict the evolution of the emission spectra as a function of pump pulse energy for these two samples, respectively. In the sub-threshold regime, the spectra exhibit weak broadband photoluminescence characteristic of the RhB-doped polymer. As the excitation energy exceeds a critical limit, sharp resonant peaks emerge, confirming the onset of stimulated oscillation in WGMs. For both environments, the full width at half maximum (FWHM) of the dominant lasing peak, displayed in the insets of \figsref{fig:lasing}(a) and \ref{fig:lasing}(b), was determined by fitting a Lorentzian profile. In air, the dominant lasing emission appeared at \SI{642.39}{\nano\metre} with an FWHM of \SI{0.52}{\nano\metre}. For the sample characterized in water, the dominant mode at \SI{625.16}{\nano\metre} exhibited an FWHM of \SI{0.74}{\nano\metre}. Using $Q = \lambda_{0}/\Delta\lambda_{\mathrm{FWHM}}$, where $\Delta\lambda_{\mathrm{FWHM}}$ represents the Lorentzian linewidth, the spectral quality factors derived from the measured emission linewidths were approximately $1.2\times10^{3}$ in air and $8.4\times10^{2}$ in water. The sample measured in water exhibited a broader emission linewidth while maintaining lasing operation. This performance substantiates the viability of the fiber-tip microring laser for robust sensing within aqueous environments.

The threshold behavior is specified by the light-in versus light-out characteristics presented in \figref{fig:lasing}(c). The transition from spontaneous to stimulated emission is marked by a distinct slope change in the integrated output energy. Based on linear fitting of the data, the lasing thresholds in air and water were extracted to be approximately \SI{0.224}{\micro\joule} and \SI{0.458}{\micro\joule}, respectively. Immersion in aqueous media reduces the RI contrast between the SU-8 polymer resonator ($n \approx 1.59$) and the surrounding environment ($n \approx 1.332$ for water at \SI{25}{\celsius}). This reduction exacerbates radiation losses and weakens mode confinement, requiring higher pump intensities to achieve the gain--loss compensation required for lasing. To determine the pump energy reaching the fiber-tip microring laser, optical losses along the excitation path, including fiber insertion loss, were considered. These losses were measured using a \SI{532}{\nano\metre} continuous-wave laser, with optical powers of \SI{150.6}{\milli\watt} and \SI{48}{\milli\watt} recorded at the laser output and fiber tip, respectively. The resulting attenuation factor of 3.14 was applied in calculating the reported lasing thresholds.

The BRIS measurements were conducted to evaluate the sensing response of the fabricated fiber-tip microlasers under controlled RI variations using aqueous glucose solutions. Prior to the BRIS measurements, the sensing chamber was filled with deionized water, and the microlaser was allowed to stabilize for approximately \SI{30}{min} to minimize transient spectral shifts associated with water uptake by the SU-8 polymer. The solutions were manually injected into the sensing chamber using syringes at a low flow rate; however, minor air-bubble formation could not be entirely avoided. Bubbles near the fiber tip or the transparent observation window may introduce measurement artifacts through local RI variations and optical scattering. Therefore, each injection was followed by a \textasciitilde\SI{3}{min} inspection period using the CCD camera to confirm the absence of bubbles. Although not implemented in this study, an automated syringe pump with an inline bubble trap could be used in future measurements to minimize bubble formation and enable uninterrupted sequential delivery of different solutions.

\begin{figure}[tb]
  \centering
  \includegraphics[width=\linewidth]{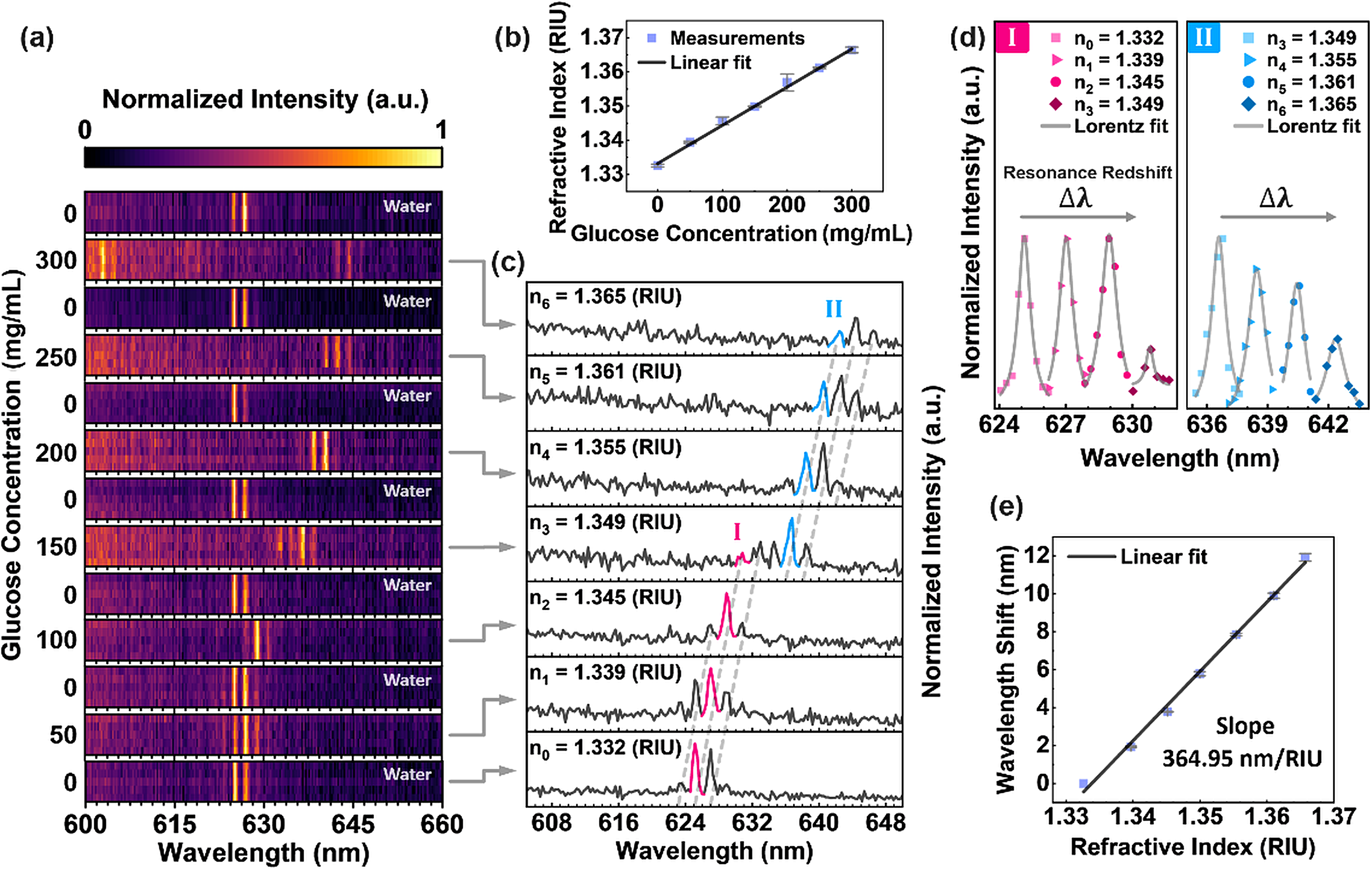}
  \caption{BRIS characterization of the fabricated fiber-tip microlaser. (a)~Normalized emission spectra for aqueous glucose solutions (\SIrange{0}{300}{\milli\gram\per\milli\litre}), with repeated water references demonstrating spectral stability and reversibility. (b)~Refractive index versus glucose concentration, with a linear fit used to estimate the RI of each tested solution. Error bars indicate the standard deviation of four refractometer measurements per concentration. (c)~Emission spectra from (a), with water-reference spectra omitted, showing RI-induced wavelength redshifts. (d)~Lasing peak tracks I and II extracted from (c), with Lorentzian fits. (e)~Wavelength shift versus surrounding refractive index, with a fitted sensitivity of \SI{364.95}{\nano\metre\per\RIU}; error bars indicate the standard deviation across tracked lasing-mode shifts from the same device.}
  \label{fig:bris}
\end{figure}

The emission spectra for each solution were recorded continuously over a \SI{10}{\second} interval. The normalized emission spectra of the microring laser in different solutions are presented in \figref{fig:bris}(a). All spectra were normalized to their respective peak intensities to highlight wavelength shifts while reducing intensity-related variations. Between successive glucose concentrations, the chamber was flushed with deionized water to remove residues from the previous glucose solution and prevent carry-over effects, ensuring that each measurement started from a clean baseline. As shown in \figref{fig:bris}(a), with increasing surrounding RI, the lasing modes exhibit an overall redshift, consistent with the increase in the effective optical path length of the WGMs. At the glucose concentration of \SI{300}{\milli\gram\per\milli\litre}, corresponding to the refractive index of 1.365, the dominant lasing peak shifts to approximately \SI{645}{\nano\metre}, and additional cavity modes emerge on the shorter-wavelength side. This behavior, which has also been observed in the simulations, does not indicate a reversal of the RI response; rather, it originates from mode competition and mode hopping within the finite gain bandwidth of the active cavity. As the primary lasing mode approaches the edge of the gain spectrum, adjacent longitudinal modes with higher net gain are preferentially amplified and dominate the emission.

Furthermore, the water reference spectra in \figref{fig:bris}(a) remained stable throughout the experiment, with a wavelength standard deviation of only \SI{0.03}{\nano\metre} around the resonance peak at \SI{625.14}{\nano\metre}. The repeated reference measurements also demonstrate the reversibility of the sensor response within the investigated range. During data acquisition, the pump power was dynamically adjusted slightly above the lasing threshold to maintain sufficient signal-to-background contrast for reliable peak tracking while minimizing dye photobleaching during prolonged excitation. The refractive index corresponding to each glucose concentration was measured using an Abbe refractometer before the sensing experiments. \figref{fig:bris}(b) shows the linear relationship between glucose concentration and refractive index, obtained from four independent measurement runs at \SI{25}{\celsius}. The resulting linear relationship was used to convert glucose concentration into refractive index in the BRIS measurements.

\figref{fig:bris}(c) illustrates the emission spectra after removal of the intermediate water reference spectra for direct comparison of the resonance shifts at different glucose concentrations. The redshift of the two representative lasing-mode tracks highlighted in \figref{fig:bris}(c) is further depicted in \figref{fig:bris}(d). For quantitative determination of the BRIS value, each lasing mode was individually fitted with a Lorentzian profile to determine its peak center, and the peak positions were subsequently tracked as the surrounding RI changed. The extracted resonance-wavelength shift as a function of the surrounding RI, as presented in \figref{fig:bris}(e), is fitted using a linear regression model with a high coefficient of determination ($R^{2} = 0.997$), indicating an approximately linear response over the investigated range. The slope of the fitted line corresponds to an experimental BRIS value of approximately $S_{\mathrm{RI}} = \SI{364.95}{\nano\metre\per\RIU}$. The limit of detection (LOD) was estimated to be \SI{2.47e-4}{\RIU} using $\mathrm{LOD} = 3\sigma/S_{\mathrm{RI}}$~\cite{White2008}, with $\sigma = \SI{0.03}{\nano\metre}$ corresponding to the standard deviation of the reference peak wavelength across repeated water measurements.

The overall redshift response and spectral behavior of the tested samples are qualitatively consistent with the trends predicted by the FDTD simulations. The small variations observed in the obtained $S_{\mathrm{RI}}$ values among different samples fall within the expected range and can be attributed to minor fabrication tolerances, natural material inhomogeneities, and slight differences in mode-excitation conditions. Importantly, the wavelength shifts induced by RI variations remained internally consistent within each sample's spectral analysis. Overall, the results indicate the robustness of bulk RI sensing based on resonance-wavelength shifts in the active WGM microring laser, exhibiting high linearity across the examined RI range. Table~\ref{tab:comparison} compares the RI sensitivity and working range of our sensor with selected previously reported passive and active WGM-based RI sensors, including their structural configurations, fabrication techniques, and coupling methods.

\begin{table}[tb]
  \centering
  \caption{Comparison of sensitivity and working range of different WGM-based refractometric sensors in passive and active operation modes.}
  \label{tab:comparison}
  \small
  \renewcommand{\arraystretch}{1.25}
  \begin{tabularx}{\linewidth}{@{}l l L >{\raggedright\arraybackslash}p{2.6cm} c c@{}}
    \toprule
    Ref. & Mode & Sensor configuration (structure / fabrication technique) & Excitation / readout & \makecell{Sensitivity\\(nm/RIU)} & \makecell{RI\\range} \\
    \midrule
    \cite{Halendy2023} & Passive & Fiber-tip polymer (IP-Dip2) microring resonator / two-photon polymerization & Fiber-tip / reflection & 25--63 & 1.00--1.35 \\
    \cite{Guo2019} & Passive & Silica optofluidic microbubble resonator / arc-discharge-assisted fabrication & Tapered fiber / transmission & 11.39 & 1.332--1.344 \\
    \cite{Zhao2026} & Passive & In-fiber liquid-filled microsphere resonator / fs-laser micromachining & In-fiber / reflection & 439 & 1.672--1.692 \\
    \cite{Krammer2017} & Active & PM597-doped PMMA microdisk laser / EBL \& XeF$_2$ etching & Free-space / free-space & 60--150 & 1.333--1.336 \\
    \cite{Kim2019} & Active & Silicon-nanocluster-embedded active microdisk resonators & Free-space / free-space & 226.7 & 1.331--1.359 \\
    \cite{Yan2021} & Active & Fiber-integrated Er-doped microsphere laser / arc discharge \& fs-laser micromachining & In-fiber / transmission & 26 & 1.34--1.375 \\
    \midrule
    This work & Active & Fiber-tip RhB-doped SU-8 microring laser / UV laser direct writing & Fiber-tip / free-space & \textbf{364.95} & 1.332--1.365 \\
    \bottomrule
  \end{tabularx}

  \vspace{2pt}
  \raggedright\footnotesize fs-laser: femtosecond laser; EBL: electron-beam lithography.
\end{table}

\section{Water-induced changes and photobleaching effects}
\label{sec:stability}

\begin{figure}[tb]
  \centering
  \includegraphics[width=\linewidth]{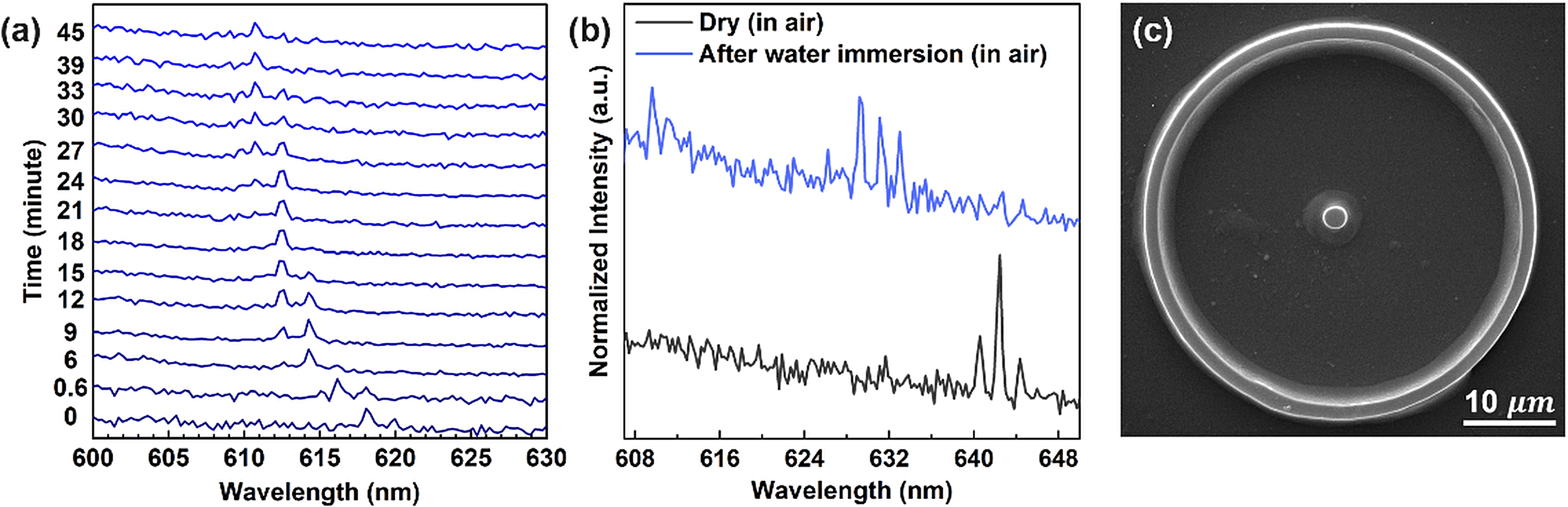}
  \caption{Temporal stability and water-induced structural effects in SU-8 fiber-tip microring lasers. (a)~Time-resolved emission spectra during continuous immersion in deionized water, showing initial wavelength drift over the first \textasciitilde\SI{10}{min}, followed by quasi-stabilization after \SIrange{25}{30}{min}. (b)~Emission spectra measured in air for an unexposed dry microlaser and for the same device after water immersion and subsequent drying, revealing persistent water-induced spectral changes. (c)~SEM image after prolonged operation in aqueous environments, showing sidewall alterations consistent with polymer swelling and structural relaxation.}
  \label{fig:water}
\end{figure}

In polymer-based resonators, water absorption and swelling can modify both the cavity geometry and the effective RI~\cite{Eryurek2017}. Such variations may result in time-dependent resonance shifts. Time-resolved measurements were performed to evaluate the spectral stability of the SU-8 fiber-tip microring laser under continuous immersion in water. The probe was placed inside a fluidic chamber filled with deionized water, and emission spectra were recorded at predefined intervals over \SI{45}{min} using intermittent optical pumping. This measurement protocol minimized continuous optical excitation, thereby reducing thermal effects and photobleaching-induced artifacts. As shown in \figref{fig:water}(a), an immediate wavelength shift is observed after water exposure. The drift is most pronounced during approximately the first \SI{10}{min}, indicating a transient response of the polymer cavity. This behavior is consistent with water diffusion into the SU-8 matrix and the associated swelling, which modifies the effective RI and optical path length of the WGMs. With increasing immersion time, the drift rate gradually decreases. After approximately \SIrange{25}{30}{min}, the lasing peaks approach quasi-stable positions. This regime persisted throughout the remaining monitoring period, supporting the use of the microlaser for bulk RI sensing after an initial equilibration period.

To assess reversibility, a microlaser previously immersed in water was dried and recharacterized in air. As illustrated in \figref{fig:water}(b), the emission spectrum of the dried device differs from that of an unexposed dry structure. This result indicates persistent water-induced changes that remain after drying. Although lasing operation is preserved, the spectral features are shifted and partially altered. The SEM image in \figref{fig:water}(c) shows sidewall irregularities that may be associated with swelling and partial structural relaxation of the SU-8 polymer. Despite these persistent water-induced changes, the stabilized microring laser remains suitable for RI sensing in aqueous environments.

\begin{figure}[tb]
  \centering
  \includegraphics[width=0.78\linewidth]{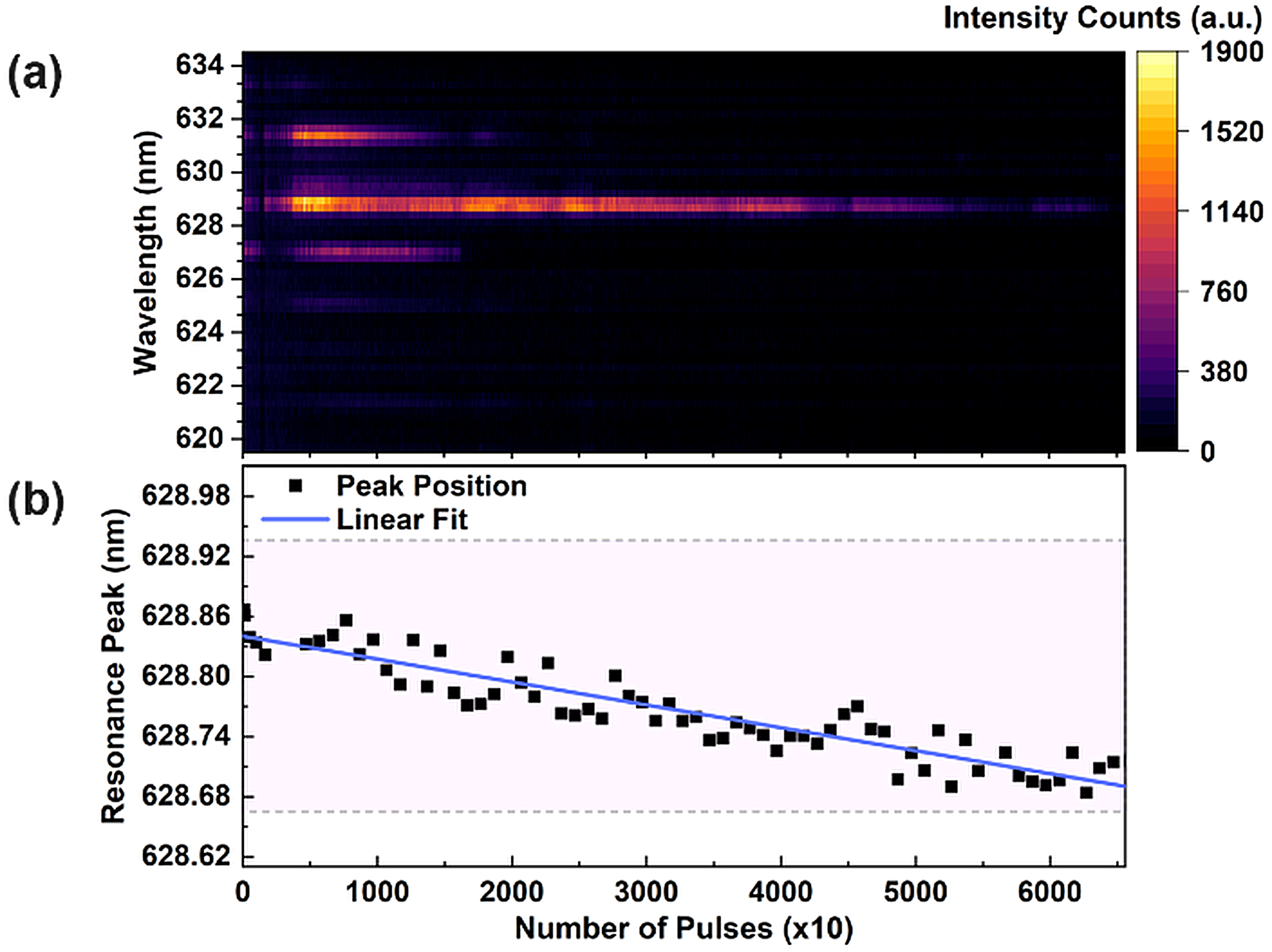}
  \caption{Photobleaching effects and operational lifetime of the fiber-tip microlaser. (a)~Time-resolved spectral heatmap recorded under continuous pulsed optical pumping, showing gradual intensity decay over \textasciitilde$6.5\times10^{4}$ pump pulses. (b)~Extracted resonance wavelength versus accumulated pump pulses with linear fitting, showing a slow photobleaching-induced blueshift below the spectrometer resolution (\textasciitilde\SI{0.28}{\nano\metre}).}
  \label{fig:bleach}
\end{figure}

Active microlasers based on organic gain media are inherently subject to photophysical degradation mechanisms such as photobleaching and fluorophore quenching under prolonged optical pumping~\cite{Vellaichamy2023}. These processes can gradually alter the gain characteristics and emission behavior of the microlaser~\cite{Krammer2017,Wondimu2018}. To investigate these effects, a laser lifetime test was conducted under continuous pulsed excitation until noticeable degradation of the emission signal was observed. The microlaser was pumped with nanosecond pulses at a repetition rate of \SI{10}{\hertz} and a pulse energy of approximately \SI{3.18}{\micro\joule}, corresponding to stable lasing conditions without immediate thermal damage. \figref{fig:bleach}(a) illustrates the temporal evolution of the microlaser emission spectra recorded over approximately $6.5\times10^{4}$ pump pulses. The gradual reduction in emission intensity shows progressive degradation of the organic gain medium and is attributed primarily to photobleaching. Lasing was nevertheless maintained throughout the measurement period, demonstrating sustained operation under the applied pumping conditions.

In addition to intensity degradation, photobleaching can result in gradual spectral shifts due to changes in the RI of the polymer--dye composite. In some previously reported organic microlasers, this effect was typically observed as a blueshift of the lasing wavelength~\cite{Krammer2017,Wondimu2018}. In the present work, however, such a blueshift is not clearly visible in \figref{fig:bleach}(a). To analyze this behavior, a quantitative peak-tracking method was applied. Approximately 75 spectra were selected from different stages of the lifetime test, and a Lorentzian fit was used to determine the central wavelength of the selected lasing peak. \figref{fig:bleach}(b) shows the extracted resonance wavelength as a function of the accumulated number of pump pulses. A gradual shift toward shorter wavelengths is observed, consistent with a photobleaching-induced blueshift. The extracted wavelength shift follows an approximately linear trend over the investigated pulse range, as indicated by linear fitting in \figref{fig:bleach}(b). The total shift remains below the instrumental spectral resolution (\textasciitilde\SI{0.28}{\nano\metre}), as marked by the dashed line and shaded region. Therefore, the blueshift is not directly distinguishable in the spectral heatmap but becomes evident through quantitative peak analysis. The observed photobleaching behavior is strongly dependent on the pump conditions. Higher pump energies can accelerate fluorophore degradation and reduce the effective laser lifetime, while lower pump energies would prolong operation at the expense of reduced signal strength. This trade-off highlights the importance of optimizing pump energy for sensing applications to balance signal stability, operational lifetime, and measurement accuracy.

\section{Conclusion and outlook}
\label{sec:conclusion}

The optical coupling and readout configuration is an important factor in the practical implementation of WGM sensors. Efficient and stable evanescent coupling requires careful control of the resonator--coupler gap and its protection against mechanical disturbances and contamination. Integrating an active WGM resonator directly on an optical-fiber facet provides a stable configuration in which this external coupling interface is avoided. Direct fabrication on a fiber tip, however, is constrained by its limited area and the geometry of the fiber, making many conventional planar microfabrication techniques difficult to implement. In this work, we addressed these optical-integration and fabrication requirements by combining micro-droplet coating with UV laser direct writing to fabricate an RhB-doped SU-8 microring laser directly on the flat fiber tip. This maskless and site-selective fabrication process enables the microring to be written approximately concentric with the fiber core. The fabricated fiber-tip microlasers exhibited stable lasing in both air and water, with lasing thresholds of \SI{0.224}{\micro\joule} and \SI{0.458}{\micro\joule}, respectively. The higher threshold in water is consistent with the reduced refractive-index contrast and increased radiation losses in the aqueous environment. The device exhibited a linear spectral response to refractive-index changes in aqueous glucose solutions, achieving a sensitivity of approximately \SI{364.95}{\nano\metre\per\RIU} and a detection limit of \SI{2.47e-4}{\RIU}.

Operation in water also revealed practical aspects of the polymer microlaser probe for aqueous sensing. Water-induced changes in the polymer produced an initial spectral drift, followed by a quasi-stable regime after approximately \SI{30}{min} of stabilization in the aqueous environment. Under prolonged optical excitation, photobleaching primarily reduced the lasing intensity, while the associated spectral shift remained below the instrumental resolution. Allowing the microlaser to stabilize in water before measurement and keeping the pump energy at a safe level to minimize photodegradation enabled stable and well-controlled operation during bulk refractive index sensitivity measurements. One further improvement could be shortening the initial stabilization period by reducing water absorption in the resonator. Optimizing the SU-8 curing conditions or employing photopolymers with lower water absorption could reduce water uptake. Water-barrier coatings may provide another route to reducing water sorption~\cite{Motevasselian2014}, although their thickness and optical properties would need to be optimized to preserve sufficient interaction between the WGM evanescent field and the surrounding analyte. Moreover, as polymers can incorporate a diverse range of gain materials, it is possible to fabricate microlasers with tailored emission and lasing characteristics~\cite{Reynolds2017,Toropov2021}. Selecting a photostable gain material could mitigate photobleaching under prolonged optical excitation and improve the long-term performance of the microlaser~\cite{Vellaichamy2023}. Appropriate gain-medium selection, together with cavity optimization, could also facilitate excitation with more compact pump sources, such as LEDs, potentially reducing the size, complexity, and cost of the overall sensing platform~\cite{Kim2019}.

An important improvement would be to replace the current free-space collection with a fully fiber-based interrogation scheme, making the device more suitable for probe-based sensing. In reflection mode, the sensing fiber could deliver the pump and collect the returning laser emission~\cite{Francois2015,Halendy2023}. Alternatively, a second fiber could be aligned with the sensing fiber for transmission-mode collection while maintaining an open sensing region for analyte access. The sensing performance could be further enhanced by integrating plasmonic nanoparticles at the microring surface. Coupling between the WGM evanescent field and localized surface plasmon resonances can enhance the optical response to analytes~\cite{Reynolds2017,Chen2021}. Functionalizing the resonator surface or the plasmonic nanoparticles with target-specific receptors could extend the fiber-tip microlaser toward selective biochemical and potentially single-molecule sensing~\cite{Toropov2021}. Plasmonic coupling can also assist the lasing characteristics; in RhB-doped SU-8 microring lasers, Au nanoparticles have been used to achieve single-mode lasing~\cite{Moradiani2023}. Aside from plasmonic modification, the Vernier effect~\cite{Jahangiri2023} and PT-symmetry breaking~\cite{Kavousi2026} have been exploited in coupled WGM resonators for mode selection and single-mode lasing, as well as for enhanced sensing. Theoretical studies have predicted improved refractive-index sensitivity near exceptional points~\cite{Chaudhary2023}. Experimentally, enhanced sensing has been demonstrated using Vernier-coupled ring lasers~\cite{Ren2012} and WGM microcavities operating near an exceptional point~\cite{Chen2017}. Coupled planar microlasers on the fiber tip could therefore be explored for manipulating the lasing characteristics and improving the sensing performance. Together, these developments could advance the proposed platform toward practical, target-specific fiber-tip microlaser probes for \emph{in situ} environmental and biochemical sensing.

\section*{Author contributions}
G.-M.~Parsanasab conceived and supervised the project. S.~M.~Hosseini and M.~Yaghoobi performed the simulations, fabrication processes, and experimental characterization of the fiber-tip integrated microring laser. G.-M.~Parsanasab and L.~Hajshahvaladi contributed to the experimental discussions, simulation evaluations, and interpretation of the results. M.~I.~Zibaii contributed to the design and implementation of the refractive index sensing experiments. L.~Hajshahvaladi and S.~M.~Hosseini prepared the manuscript and generated the figures and schematic diagrams, with input from all authors. All authors discussed the results and contributed to the final version of the manuscript.

\section*{Disclosures}
The authors declare no conflicts of interest.

\section*{Data availability}
Data underlying the results presented in this paper are not publicly available at this time but may be obtained from the authors upon reasonable request.

\bibliographystyle{unsrtnat}
\bibliography{references}

\end{document}